\documentclass[conference]{IEEEtran}

\IEEEoverridecommandlockouts 

\ifCLASSINFOpdf

\else
\fi
\usepackage{mathtools}
\usepackage{amsmath}
\usepackage{amssymb}
\usepackage{cite}
\usepackage{graphicx}
\usepackage{multirow}
\usepackage{siunitx}
\usepackage{multirow}
\usepackage{siunitx}
\usepackage{graphicx}
\usepackage{graphicx}
\usepackage{algorithm, algorithmic}
\usepackage{color}
\usepackage{xcolor}
\DeclareMathOperator{\E}{\mathbb{E}}
\usepackage{geometry}
\begin{document}
\title{A Low-Complexity Recursive Approach Toward Code-Domain NOMA for Massive Communications} 

 \author{%
   \IEEEauthorblockN{Mohammad Vahid Jamali,~and Hessam Mahdavifar}
   \IEEEauthorblockA{Electrical Engineering and Computer Science Department, University of Michigan, Ann Arbor, MI 48109, USA\\
                     Email: mvjamali@umich.edu, hessam@umich.edu}
 \vspace{-0.27in}
}

\maketitle
\begin{abstract}
Nonorthogonal multiple access (NOMA) is a promising technology to meet the demands of the next generation wireless networks on massive connectivity, high throughput and reliability, improved fairness, and low latency. In this context, code-domain NOMA which attempts to serve $K$ users in $M\leq K$ orthogonal resource blocks, using a pattern matrix, is of utmost interest. However, extending the pattern matrix dimensions severely increases the detection complexity and hampers on the significant advantages that can be achieved using large pattern matrices. In this paper, we propose a novel approach toward code-domain NOMA which factorizes the pattern matrix as the Kronecker product of some other factor matrices each with a smaller dimension. Therefore, both the pattern matrix design at the transmitter side and the mixed symbols' detection at the receiver side can be performed over much smaller dimensions and with a remarkably reduced complexity and latency. As a consequence, the system can significantly be overloaded to effectively support the requirements of the next generation wireless networks without any considerable increase on the system complexity.
\end{abstract}


\section{Introduction}
Nonorthogonal multiple access (NOMA) is becoming one of the key enabling technologies for the fifth-generation (5G) wireless networks and the applications beyond, including the Internet of Things (IoT), and massive machine-type communications (MMTC) \cite{dai2015non,chen2018toward}. NOMA, through serving multiple users in the same orthogonal resource block, or equivalently, a set of users in a smaller set of resource elements (REs), attempts to significantly increase the system throughput and reliability, improve the users' fairness, reduce the latency, and support the massive connectivity \cite{ding2017survey}. NOMA is a general principle and many of the recently proposed multiple access techniques can be considered as special cases. However, generally speaking, NOMA techniques can be classified into two different categories: power-domain NOMA, and code-domain NOMA.

Power-domain NOMA serves multiple users in the same orthogonal resource block through allocating different power levels to different users. In contrast to the traditional power allocation methods, the users with poorer channel conditions are assigned more powers \cite{ding2014performance}. Power-domain NOMA has attracted significant attention in recent years and many interesting problems have been addressed in this context, including cooperative NOMA \cite{ding2015cooperative}, NOMA with simultaneous wireless information and power transfer (SWIPT) \cite{liu2016cooperative}, {multiple-input multiple-output (MIMO) NOMA \cite{sun2015ergodic}, and NOMA in millimeter wave communications \cite{ding2017random}}. Detection procedure in power-domain NOMA highly relies on successive interference cancellation (SIC). Therefore, when the channel condition of different users paired together is close to each other, SIC-based power-domain NOMA does not work well. In order to enable (near) power-balanced NOMA, the authors of \cite{NCMA} proposed network-coded multiple access (NOMA); a new NOMA architecture which jointly employs the physical-layer network coding and multiuser decoding to significantly increase the NOMA throughput.

Code-domain NOMA, on the other hand, attempts to serve a set of users, say $K$, in a set of $M$ orthogonal resource blocks, $M\leq K$, using a pattern matrix. The pattern matrix comprises $K$ pattern vectors assigned to each user specifying the set of REs available to them. As an interesting feature of code-domain NOMA, it can work well even in power-balanced scenarios when each user is served by a unique pattern vector. Nevertheless, code-domain NOMA has received relatively less attention compared to the power-domain counterpart, because it usually requires rather complex multiuser nonlinear detection algorithms, such as massage passing algorithm (MPA), maximum likelihood (ML) detection, and maximum  \textit{a posteriori} (MAP) detection. 
In this context, sparse code multiple access (SCMA) was proposed in \cite{nikopour2013sparse} based on directly mapping the incoming bits to multidimensional codewords of a SCMA codebook set. {A low complexity SCMA decoding algorithm has been further suggested in \cite{wei2016low} based on list sphere
		decoding. Moreover, lattice partition multiple access (LPMA) has been proposed in 
		\cite{fang2016lattice,qiu2018lattice}
		based on multilevel lattice codes for multiple users. Interleave-grid multiple access (IGMA) has also been proposed in order to increase the user multiplexing capability and improve the performance \cite{xiong2017advanced}.}
		 Recently, pattern division multiple access (PDMA) has been introduced in \cite{chen2017pattern}, where the pattern vectors are designed with disparate order of transmission diversity to mitigate the error propagation problem in SIC receivers. As a consequence, the sparsity feature of the pattern matrix does not necessarily hold, i.e., the number of REs allocated to a particular user may be comparable to the total number of resource blocks $M$.

Design of the pattern matrix plays a critical role in code-domain NOMA to balance the trade-off between the system performance and complexity. Impact of pattern matrix on the average sum rate of SCMA systems has been investigated in \cite{yang2016impact} where a low-complexity iterative algorithm has been proposed to facilitate the design of pattern matrix. Moreover, the total throughput of low-density code-domain (LDCD) NOMA has recently been characterized in \cite{shental2017low} for regular random pattern matrices with large dimensions. It is well understood that, for a given overload factor $\beta=K/M$, the higher the dimension of pattern matrix we have, the better the performance we get. This is because more number of REs can be assigned to each user for a higher dimension of pattern matrix \cite{chen2017pattern}. However, increasing the pattern matrix dimension results in a significantly higher detection complexity. Inspired by this trade-off between the system performance and detection complexity, we propose a novel approach toward code-domain NOMA which factorizes the pattern matrix as the Kronecker product of some other smaller factor matrices. As a consequence, both the pattern matrix design at the transmitter and the mixed symbols' detection at the receiver can be performed over much smaller dimensions and with a remarkably reduced complexity, through recursive detection; hence, the system performance can significantly be improved without any considerable increase on the complexity.

The rest of the paper is organized as follows. In Section II, we briefly review the basics of code-domain NOMA with pattern matrix and highlight the preliminaries. In Section III, we describe the proposed method, and provide an illustrating example. In Section IV, we classify the design procedures, obtain the sum rate formula, and analyze the detection complexity. We provide the numerical results in Section V, and conclude the paper in Section VI. 
\section{Preliminaries and System Model}
We consider a collection of $K$ users communicating using $M$ REs, and define the overload factor as $\beta=K/M\geq 1$ implying the multiplexing gain of PDMA compared to orthogonal multiple access (OMA).
 The modulation symbol  $x_k$ of the $k$-th user is spread over $M$ orthogonal REs using the pattern vector $\boldsymbol{g}_k$ as $\boldsymbol{v}_k=\boldsymbol{g}_k x_k$, $1\leq k\leq K$, where $\boldsymbol{g}_k$ is an $M\times 1$ binary vector defining the set of REs available to the $k$-th user; the $k$-th user can use the $i$-th RE, $1\leq i\leq M$, if the $i$-th element of $\boldsymbol{g}_k$ is ``$1$'', i.e., ${g}_{i,k}=\boldsymbol{g}_{k}(i)=1$, and vice versa. Then the ($M\times K$)-dimensional pattern matrix can be constructed as \cite{chen2017pattern}
\begin{align}\label{eq1}
\boldsymbol{G}_{M\times K}=
\begin{bmatrix}
\boldsymbol{g}_{1} & \boldsymbol{g}_{2}& \cdots &\boldsymbol{g}_{K}
\end{bmatrix}=
\begin{bmatrix}
{g}_{i,k}
\end{bmatrix}_{{M\times K}}.
\end{align}

For the uplink transmission each $k$-th user transmits the spread symbol, using its pattern vector, to the base station (BS). Therefore, assuming perfect synchronization at the BS, the received ($M\times 1$)-dimensional vector, comprising the received signal at each of $M$ REs, can be modeled as the Gaussian vector channel model of
\begin{align}\label{eq2}
\boldsymbol{y}=\sum\nolimits_{k=1}^{K}{\rm diag}(\boldsymbol{h}_k)\boldsymbol{v}_k+\boldsymbol{n},
\end{align}
where $\boldsymbol{h}_k$ is the vector modeling the uplink channel response of the $k$-th user at all of $M$ REs, and $\boldsymbol{n}$ is the noise vector at the BS with length $M$. Furthermore, ${\rm diag}(\boldsymbol{h}_k)$ is a diagonal matrix consisting of the elements of $\boldsymbol{h}_k$. Then the uplink transmission model can be reformulated as
\begin{align}\label{eq3}
\boldsymbol{y}=\boldsymbol{H}\boldsymbol{x}+\boldsymbol{n},
\end{align}
in which $\boldsymbol{x}=[x_1,x_2,...,x_K]^T$, $\boldsymbol{H}=\boldsymbol{\mathcal{H}}{\bullet}\boldsymbol{G}_{M\times K}$ is the PDMA equivalent uplink channel response, $\boldsymbol{\mathcal{H}}=[\boldsymbol{h}_1,\boldsymbol{h}_2,...,\boldsymbol{h}_K]$, and $\bullet$ denotes the element-wise product \cite{chen2017pattern}.

Moreover, for the downlink transmission the BS first encodes the data symbol of each user according to its pattern vector and then transmits the superimposed encoded symbols $\sum_{j=1}^{K}\boldsymbol{v}_j$ through the channel. Therefore, the received signal at the $k$-th user can be expressed as
\begin{align}\label{eq4}
\boldsymbol{y}_k&={\rm diag}(\boldsymbol{h}_k)\sum\nolimits_{j=1}^{K}\boldsymbol{g}_jx_j+\boldsymbol{n}_k
=\boldsymbol{H}_k\boldsymbol{x}+\boldsymbol{n}_k,
\end{align}
where $\boldsymbol{H}_k={\rm diag}(\boldsymbol{h}_k)\boldsymbol{G}_{M\times K}$ is the PDMA equivalent downlink channel response of the $k$-th user. Moreover, $\boldsymbol{h}_k$ and $\boldsymbol{n}_k$ are, respectively, the downlink channel response and noise vector at the $k$-th user with length $M$ \cite{chen2017pattern}.
{We further assume that the channel state information (CSI) is available at the transmitter side such that the channel gains are equalized from the receiver viewpoint. Therefore, the received signal vector for both uplink and downlink transmissions can be expressed as
$\boldsymbol{y}=\boldsymbol{G}\boldsymbol{x}+\boldsymbol{n}$.}

As a rather simple multiuser detection algorithm in NOMA systems, SIC properly trades off between the system performance and complexity. However, SIC receivers suffer from error propagation problem meaning that the system performance highly relies on the correctness of early-detected users. To deal with this problem, we can either improve the reliability of the firstly-decoded users or employ more advanced detection algorithms such as ML and MAP. In \cite{chen2017pattern}, disparate diversity orders are adopted for different users by assigning patterns with heavier weights to those early-detected users in order to increase their transmission reliability. Furthermore, employing more advanced detection algorithms severely increases the system complexity especially for larger dimension of the pattern matrix; this may hinder their practical implementation particularly for downlink transmission where the users are supposed to have a lower computational resources than the BS. In the next section, we elaborate how our proposed method can support much larger dimensions of the pattern matrices, even with heavier pattern weights, to boost massive connectivity without a significant increase on the overall system complexity.

\section{Proposed Design Method}
\subsection{Fundamentals of the Recursive Pattern Matrix Design}
Depending on the number of users and the available REs to them, we propose to factorize the
  pattern matrix as
\begin{align}\label{eq5}
\boldsymbol{G}_{M\times K}=\boldsymbol{G}^{(1)}_{m_1\times k_1}\otimes \boldsymbol{G}^{(2)}_{m_2\times k_2} \otimes ... \otimes \boldsymbol{G}^{(L)}_{m_L\times k_L},
\end{align}
where $\otimes$ denotes the Kronecker product defined as
\begin{align}\label{eq6}
\boldsymbol{A}_{m\times k}\otimes \boldsymbol{B}_{m'\times k'}=\begin{bmatrix}
a_{11}\boldsymbol{B} & \cdots & a_{1k}\boldsymbol{B} \\ 
 \vdots & \ddots  & \vdots \\ 
 a_{m1}\boldsymbol{B} & \cdots  & a_{mk}\boldsymbol{B}
\end{bmatrix}
\end{align} 
for
 any two matrices $\boldsymbol{A}$ and $\boldsymbol{B}$. Then the dimensions of the resulting pattern matrix in \eqref{eq5} can be obtained based on the dimensions of the factor matrices as $M=\prod_{l=1}^{L}m_l$, and $K=\prod_{l=1}^{L}k_l$. In general, if at least one of $M$ or $K$ is not a prime number, we can find some $m_l>1$ or $k_l>1$ to factorize the pattern matrix as the Kronecker product of some smaller factor matrices. On the other hand, if both $M$ and $K$ are prime numbers, $L$ is equal to one and the design procedure simplifies to that of  conventional pattern matrix design (such as PDMA \cite{chen2017pattern}). Note that both $M$ and $K$ are design parameters and we can always group an appropriate number of users with a specific number of REs to optimize the system in terms of performance and complexity. 

The proposed structure not only alleviates the detection complexity at the receiver side (through recursive detection as will be clarified through an illustrating example in the next subsection) but also significantly reduces the search domain at the transmitting party, enabling the usage of large pattern matrices with a reasonable complexity for massive communications. In fact, a regular pattern matrix design requires a comprehensive search over all $\binom{{2^M}-1}{K}$ possible ($M\times K$)-dimensional matrices with distinct nonzero columns (patterns assigned to each user) to find an optimal pattern matrix \cite{chen2017pattern}. On the other hand, it is easy to show that satisfying distinct nonzero columns for the overall pattern matrix $\boldsymbol{G}_{M\times K}$ of the form \eqref{eq5} requires distinct nonzero columns for all of the factor matrices $\boldsymbol{G}^{(l)}_{m_l\times k_l}$, $l=1,2,...,L$. In this case, the total search domain for our proposed design method reduces to $\prod_{l=1}^{L}\binom{2^{m_l}-1}{k_l}$ which is much smaller than $\binom{{2^M}-1}{K}$. For example, for $M=6$ and $K=9$, the regular pattern matrix design requires searching over $\binom{{2^6}-1}{9}=2.36\times 10^{10}$ possible matrices, while our design method with factorizing the overall pattern matrix as the Kronecker product of $\boldsymbol{G}^{(1)}_{2\times 3} \otimes \boldsymbol{G}^{(2)}_{3\times 3}$ only needs to search over $\binom{{2^2}-1}{3}.\binom{{2^3}-1}{3}=35$ matrices. Note that if, for some $l$, $2^{m_l}-1<k_l$, then we cannot satisfy distinct nonzero columns for the factor matrix $\boldsymbol{G}^{(l)}_{m_l\times k_l}$. Consequently, as it will be demonstrated in the following example, the overall pattern matrix $\boldsymbol{G}_{M\times K}$ will have some repeated columns, i.e., the same pattern vectors for some of the users.
{In such circumstances, we can either use more advanced detection methods such as MAP, though over much smaller dimensions, or assign different power coefficients to the users having the same pattern vectors.
}
It is worth mentioning that recursive construction of large matrices based on the Kronecker product of some smaller matrices has been used in different contexts such as polar coding \cite{arikan2009channel} and its later versions like compound polar coding \cite{mahdavifar2013compound}.
\subsection{An Illustrating Example}
Here we provide an illustrating example to clarify the recursive detection method and facilitate the performance characterization procedures in the next section. 

\noindent{\textbf{Example 1.}} Consider a code-domain NOMA network with $K=18$ users over $M=9$ REs realized using the Kronecker product of the following three factor matrices
 \vspace{-0.05in}
\begin{align}\label{eq7}
\boldsymbol{G}^{(1)}_{1\times 2}=\begin{bmatrix}
1 & 1
\end{bmatrix},~~~
\boldsymbol{G}^{(2)}_{3\times 3}=\boldsymbol{G}^{(3)}_{3\times 3}=
\begin{bmatrix}
1 & 1 & 0 \\ 
1 & 0 & 1\\ 
 0 & 1  & 1
\end{bmatrix}.
\end{align}
Then the resulting overall pattern matrix can be obtained as $\boldsymbol{G}_{9\times 18}=\boldsymbol{G}^{(1)}_{1\times 2}\otimes \boldsymbol{G}^{(2)}_{3\times 3} \otimes \boldsymbol{G}^{(3)}_{3\times 3}$. And assuming the availability of CSI, the received signal vector, either in uplink or downlink reception, can be represented as $\boldsymbol{y}_{9\times 1}=\boldsymbol{G}_{9\times 18}\boldsymbol{x}_{18\times 1}+\boldsymbol{n}_{9\times 1}$.

It is easy to verify that the $k'$-th and ($k'+9$)-th columns, $k'=1,2,...,9$, of the overall pattern matrix $\boldsymbol{G}_{9\times 18}$ are the same, meaning that both the $k'$-th and ($k'+9$)-th users have the same pattern vector $\boldsymbol{g}_{k'}$. This is because both of the elements of $\boldsymbol{G}^{(1)}_{1\times 2}$ are the same and this duplicates $\boldsymbol{G}^{(2)}_{3\times 3}\otimes \boldsymbol{G}^{(3)}_{3\times 3}$ in the construction of the overall pattern matrix. As a result, the data symbols of the $k'$-th and ($k'+9$)-th users, i.e., $x_{k'}$ and $x_{k'+9}$, are always coupled together, leading us to define new symbols $t_{k'}=x_{k'}+x_{k'+9}$. Therefore, we can write the received signals at different REs $y_i$'s, $i=1,2,...,9$, as the linear combination of $t_{k'}$'s, defined based on the rows of the square matrix $\boldsymbol{G}^{(2)}_{3\times 3}\otimes \boldsymbol{G}^{(3)}_{3\times 3}$, plus the noise components $n_i$'s {(e.g., the received signal at the first RE is given by $y_1=x_1+x_2+x_4+x_5+x_{10}+x_{11}+x_{13}+x_{14}+n_1=t_1+t_2+t_4+t_5+n_1$)}.
The resulting set of $9$ equations for $y_i$'s can be analyzed using different multiuser detection methods to find $t_{k'}$'s. However, the proposed design method in this paper can recursively find the data symbols with much lower complexity. Here, for the sake of brevity, we focus on the detection of $t_1$, $t_4$, and $t_7$. {The rest of the symbols can be detected in a similar recursive procedure.}
 Let us define new variables $Z_1$, $Z_4$, and $Z_7$ as the combination of some of the received signals as
\begin{align}\label{eq9}
Z_1&=y_1+y_2-y_3=2t_1+2t_4+n'_1,\nonumber\\
Z_4&=y_4+y_5-y_6=2t_1+2t_7+n'_4,\nonumber\\
Z_7&=y_7+y_8-y_9=2t_4+2t_7+n'_7,
\end{align}
where $n'_1=n_1+n_2-n_3$, $n'_4=n_4+n_5-n_6$, and $n'_7=n_7+n_8-n_9$. Therefore, the effective signal-to-noise ratio (SNR) increased by a factor of $4/3$ because the noise elements in different REs $n_{k'}$'s are independent, i.e., each of $n'_{k'}$'s has the variance of $\sigma'^2=3\sigma^2$, where $\sigma^2$ is the variance of original noise elements $n_{k'}$'s.
Again, we can apply the recursive procedure to get much simpler equations as
\begin{align}\label{eq10}
U_1&=Z_1+Z_4-Z_7=4t_1+n''_1,\nonumber\\
U_4&=Z_1+Z_7-Z_4=4t_4+n''_4,\nonumber\\
U_7&=Z_4+Z_7-Z_1=4t_7+n''_7,
\end{align}
in which $n''_1=n'_1+n'_4-n'_7$, $n''_4=n'_1+n'_7-n'_4$, and $n''_7=n'_4+n'_7-n'_1$. Therefore, the effective SNR is increased again by a factor of $4/3$ as the resulting noise components $n''_{k'}$'s have the variance of $\sigma''^2=3\sigma'^2=3^2\sigma^2$. Finally, based on \eqref{eq10}, the original data symbols $x_k$'s can be obtained using optimal MAP detection over simple equations of the form 
\begin{align}\label{eq11}
U_{k'}=4x_{k'}+4x_{k'+9}+n''_{k'},~~k'=1,2,...,9
\end{align}
which have much smaller dimensions (i.e., $1\times 2$) compared to the original pattern matrix of dimension $9\times 18$. 

{\noindent{\textbf{Remark 1.} Although the effective reduction on the dimension of equations, using the proposed recursive detection, enables the application of advanced detection methods such as MAP, we can also exploit the power-domain NOMA concept to group the users having the same pattern vector (here the $k'$-th and ($k'+9$)-th users) in order to differentiate between those users and  further boost the performance and reduce the complexity of the system; this interesting pronlem will be explored in our future work.}}
\section{Performance Characterization}
The previous example gives a good idea about the potentials of the proposed code-domain NOMA approach and illustrates the process of the low-complexity recursive detection. However, it raises a couple of important questions which should be carefully addressed. In particular, what kind of matrices should be selected and how we can find the optimal factor matrices? How the received signals in different REs can be combined to get smaller dimensions and simpler sets of equations (e.g., converting \eqref{eq9} to \eqref{eq10})? What exactly will be the gain of such a combining in terms of increasing the effective SNR? How many equations and with what dimensions will be left at the end to perform the advanced detection algorithms, such as MAP, and then what would be the overall  detection complexity? And, given a set of factor matrices to construct a pattern matrix, what is the overall system performance in terms of the average sum rate? In this section, in order to appropriately answer these issues, we consider a special case where $\boldsymbol{G}^{(1)}_{m_1 \times k_1}=\boldsymbol{F}_{m_f \times k_f}$ is a rectangular matrix with $m_f<k_f$ and all of the other $r=L-1$  factor matrices are the same square matrices as $\boldsymbol{G}^{(l)}_{m_l \times k_l}=\boldsymbol{P}_{m_p \times m_p}$, $l=2,3,...,L$ {(hence, the overload factor is $\beta=k_f/m_f>1$)}, 
 and leave the general study on the pattern matrix construction to the future research. Therefore, the overall pattern matrix can be represented as $\boldsymbol{G}_{M \times K}=\boldsymbol{F}_{m_f \times k_f}\otimes \boldsymbol{P}^{\otimes r}_{m_p \times m_p}$, where $\boldsymbol{P}^{\otimes r}$ implies the $r$-times Kronecker product of $\boldsymbol{P}$ with itself.
\subsection{Pattern Matrix Design and Recursive Combining}
With the considered pattern matrix structure, $\boldsymbol{G}=\boldsymbol{F}\otimes \boldsymbol{P}^{\otimes r}$, both the combining procedure of the received signals in different REs and the resulting gains in the average SNRs, in each recursion, directly relate to the square factor matrix $\boldsymbol{P}$. Given the $v$-th possible factor matrix $\boldsymbol{P}_v$, $v=1,2,...,\binom{2^{m_p}-1}{m_p}$, at each $l$-th iteration, we have some set of auxiliary equations of the form
\begin{align}\label{eq13}
\!\!\!\!\!\!\begin{bmatrix}
y^l_1\\
y^l_2\\
\vdots\\
y^l_{m_p}
\end{bmatrix}\!\!=\!\!
\begin{bmatrix}
p^v_{11} & p^v_{12} & \cdots & p^v_{1m_p}\\
p^v_{21} & p^v_{22} & \cdots & p^v_{2m_p}\\
\vdots & \vdots & \ddots & \vdots\\
p^v_{m_p1} & p^v_{m_p2} & \cdots & p^v_{m_pm_p}
\end{bmatrix}\!\!
\begin{bmatrix}
x^l_1\\
x^l_2\\
\vdots\\
x^l_{m_p}
\end{bmatrix}
\!\!+\!\!
\begin{bmatrix}
n^l_1\\
n^l_2\\
\vdots\\
n^l_{m_p}
\end{bmatrix}\!,\!
\end{align}
where $p^v_{ij}$ is the $(i,j)$-th element of the factor matrix $\boldsymbol{P}_v$. And $y^l_j$, $x^l_j$, and $n^l_j$ are respectively the $j$-th auxiliary received signal, data symbol, and noise component at the $l$-th recursion {(see, e.g., the right hand side of \eqref{eq9} and recall that, with some changes of notation,  $Z_j$, $2t_j$, and $n'_j$ are equivalent to $y^l_j$, $x^l_j$, and $n^l_j$ in the notation of Eq. \eqref{eq13})}.

In order to obtain the effective combining coefficients {(such as combining $Z_{j}$'s in \eqref{eq10})} that facilitate the recursive detection with a low complexity and in the meantime provide the maximum increase of the SNRs, we find the so-called systematic representation of $\boldsymbol{P}_v$, i.e., for each $j$-th data symbol we find all the linear combinations of the rows of the factor matrix that result in the form $x^l_j\sum_{i=1}^{m_p}\alpha^{(v,j)}_ip^v_{ij}$ which only contains $x^l_j$. In this representation, $\alpha^{(v,j)}_i\in\{-1,0,1\}$, $i=1,2,...,m_p$, are the combining coefficients. Such a combination, due to the independency of the noise components $n^l_j$'s, multiplies the SNR of the previous iteration by $\gamma_{v,j}=w_{v,j}^2/\sum_{i=1}^{m_p}(\alpha^{(v,j)}_i)^2$, where $w_{v,j}=\sum_{i=1}^{m_p}\alpha^{(v,j)}_ip^v_{ij}$ is the weight of the desired data symbol $x^l_j$ after the combination. Then the selection procedure for the combining coefficients that yield the maximum SNR for each auxiliary data symbol can be summarized as Algorithm 1.

\begin{algorithm}[t]
\caption{Calculation of the combining coefficients and the corresponding individual SNRs}
\begin{algorithmic}[1]
    \STATE Input: dimension of the square matrix $m_p$
    \STATE Output: combining coefficients and SNRs for all $v$'s and $j$'s
    \FOR {$v = 1:\binom{2^{m_p}-1}{m_p}$}
    \STATE $\boldsymbol{P}_v={\begin{bmatrix}p^v_{ij}\end{bmatrix}}_{m_p\times m_p}$
        \FOR {$j = 1:m_p$}
        \STATE {\textbf{Find}} all sets of coefficients  $\left\{\alpha'^{(v,j)}_{i=1:m_p}\right\}$ {\textbf{Such That:}}
        \STATE ~~~~{\textbf{C1:}} $\alpha'^{(v,j)}_i\in\{-1,0,1\}$
                \STATE ~~~~{\textbf{C2:}} $\sum_{i=1}^{m_p}\alpha'^{(v,j)}_ip^v_{ij}={w'}_{v,j}\neq 0$
                        \STATE ~~~~{\textbf{C3:}} $\sum_{i=1}^{m_p}\alpha'^{(v,j)}_ip^v_{ij'}=0,~j'\neq j=1,2,...,m_p$
                  \STATE {\textbf{Calculate}} $\gamma'_{v,j}\!=\!\left[{w'}_{v,j}\right]^2\!\!\!{\Big/}\!\sum_{i=1}^{m_p}(\alpha'^{(v,j)}_i)^2$
    \STATE {\textbf{Output}} $\left\{\alpha^{(v,j)}_{i=1:m_p}\right\}={\rm argmax}~~ \gamma'_{v,j}$ over $\left\{\alpha'^{(v,j)}_{i=1:m_p}\right\}$
    \STATE {\textbf{Output}} $\gamma_{v,j}=\left[\sum_{i=1}^{m_p}\alpha^{(v,j)}_ip^v_{ij}\right]^2\!\!\!{\Big/}\!\sum_{i=1}^{m_p}(\alpha^{(v,j)}_i)^2$
    \ENDFOR
        \ENDFOR
\end{algorithmic}
\end{algorithm}

 
\noindent{\textbf{Remark 2.}} The optimization procedure in Algorithm 1 is performed over all $\binom{2^{m_p}-1}{m_p}$ possible factor matrices with distinct nonzero columns, and then the optimal square factor matrix, the gains on the SNRs, and the efficient combining coefficients are obtained by choosing the best answer set satisfying further constraints, e.g., maximizing the average sum rate. {We should emphasize based on our hierarchical construction that even for the  pattern matrices with large overall dimensions the search process in Algorithm 1 is performed over square factor matrices $\boldsymbol{P}$ with much smaller dimensions $m_p$; this further paves the way toward massive code-domain NOMA.}

\noindent{\textbf{Remark 3.}} {It is easy to verify that} for the factor matrices with linearly-independent rows there is only one set of coefficients
 satisfying the constraints in Algorithm 1 for each data symbol.
 
{
\subsection{Detection Procedure and its Complexity}
In this subsection, we first elaborate the proposed recursive detection method and then characterize its complexity.
\subsubsection{Detection Procedure} Assume that the communication protocol is established between the transmitters and receivers, i.e., the design parameters such as the optimal square factor matrix $\boldsymbol{P}_{m_p\times m_p}$ and combining coefficients from Algorithm 1 (also recall Remark 2), number of recursions $r=L-1$, and the rectangular matrix $\boldsymbol{F}_{m_f\times k_f}$ are known for the receiver (either a typical user during the downlink phase or the BS during the uplink transmission); therefore, the pattern matrix structure $\boldsymbol{G}_{M \times K}=\boldsymbol{F}_{m_f \times k_f}\otimes \boldsymbol{P}^{\otimes r}_{m_p \times m_p}$ is known. Given the combining coefficients, we can further define an $m_p\times m_p$ matrix $\boldsymbol{\alpha}$ comprising the combining vectors $\big\{\alpha^{(j)}_i\big\}_{i=1}^{m_p}$ as its $j$-th row such that the matrix product of $\boldsymbol{\alpha}\boldsymbol{P}$ gives an $m_p\times m_p$ diagonal matrix with the $j$-th diagonal element $w_{j}=\sum_{i=1}^{m_p}\alpha^{(j)}_ip_{ij}$ being the weight of the $j$-th data symbol after the combination (recall the constraints for the combining coefficients from Algorithm 1).

\textbf{First Recursion:} In order to perform the detection, the receiver in the first recursion takes the vector of received signals over $M_1=M=m_fm_p^r$ REs, i.e., $y_i$'s, $i=1,2,...,M$, and divides them into $M_1/m_p=m_fm_p^{r-1}$ groups of $m_p$ equations; therefore, the $i_1$-th group
includes $\{y_{(i_1-1)m_p+1},y_{(i_1-1)m_p+2},...,y_{i_1m_p}\}$. Each of these $y_i$'s contains at most $K$ different transmitted symbols $x_k$'s, $k=1,2,...,K$, since each one is constructed as a linear combination of $x_k$'s, defined based on the $i$-th row of the overall pattern matrix $\boldsymbol{G}_{M \times K}$, and the noise component $n_i$ over the $i$-th RE.
The receiver then combines the elements of each group using the combining matrix $\boldsymbol{\alpha}$ to form new $m_p$ symbols at each group; the first new symbol is constructed by combining the previous symbols using the first row of $\boldsymbol{\alpha}$ and so on. Note that such a combination can be expressed in a matrix form through multiplying $\boldsymbol{\alpha}$ by a new matrix comprising the $m_p$ rows of $\boldsymbol{G}$ corresponding to the symbols of each group. Then it is easy to verify that the $m'$-th new equation of each group (which is constructed through the $m'$-th row of $\boldsymbol{\alpha}$), $m'=1,2,...,m_p$, contains at most $K_1=K/m_p=k_fm_p^{r-1}$ different symbols from the set $\{x_{m'},x_{m'+m_p},...,x_{m'+k_fm_p^{r}-m_p}\}$ (recall that the product of $\boldsymbol{\alpha}\boldsymbol{P}$ is a diagonal matrix), i.e., the number of unknown variables is reduced by a factor of $m_p$. Also the effective SNR of each symbol in the $m'$-th new equation is increased by a factor of $\gamma_{m'}$.

\textbf{Second Recursion:} Note based on the above explanation that after the first recursion the $m'$-th new equation of each group can only contain symbols from the set $\{x_{m'},x_{m'+m_p},...,x_{m'+k_fm_p^{r}-m_p}\}$. This means different equations of a given group contain disjoint sets of symbols while equations with the same index of different groups (e.g., the $m'$-th equation of all groups) contain symbols from the same set. Therefore, the receiver in the second recursion forms $m_p$ super-groups of $M_2=M_1/m_p=m_fm_p^{r-1}$ equations (note that in the first recursion we had $m_p^0=1$ super-group of size $M_1=M$) by placing the $m'$-th new equation of each of those $m_fm_p^{r-1}$ groups in the first recursion into the $m'$-th super-group. Now, the receiver follows exactly the same procedure as the first recursion over each of these disjoint $m_p$ super-groups with the $m_p$ times smaller size of $M_2$, i.e., divides the equations in each of super-groups into $M_2/m_p=m_fm_p^{r-2}$ groups of $m_p$ equations and combines the signals within each group using the combining matrix $\boldsymbol{\alpha}$. Following the same logics we can argue that the maximum number of unknown variables at each of the new equations is reduced by a factor of $m_p$ from $K_1$ to $K_2=K_1/m_p=k_fm_p^{r-2}$, and the SNR of the symbols in the $m''$-th new equation, $m''=1,2,...,m_p$, of each of the groups in the $m'$-th super-group is increased by a factor of $\gamma_{m''}$ from $\gamma_{m'}$ in the first recursion to $\gamma_{m'}\gamma_{m''}$.

\textbf{Final Recursion:} By induction, it is easy to verify that in the $r$-th recursion we will have $m_p^{r-1}$ super-groups of size $M_r=M/m_p^{r-1}=m_fm_p$. Therefore, the receiver in the $r$-th iteration divides the equations within each of these $M_r$ super-groups into $M_r/m_p=m_f$ groups of size $m_p$ and combines the symbols of each group using $\boldsymbol{\alpha}$ to get equations containing at most $K_r=K/m_p^r=k_f$ unknown variables. Based on these arguments and following the same logics we can conclude that after the $r$-th iteration we end up with $m_p^r$ sets of $m_f$ equations each containing at most $K_r=k_f$ unknown variables which are defined based on the form of the rectangular matrix $\boldsymbol{F}$ and can be processed using advanced multiuser detection algorithms such as MAP though over the much smaller dimension of $m_f\times k_f$ instead of $M\times K$. The recursive detection procedure is
schematically shown through a tree diagram in Fig. 1 which further helps to understand the progressive development of the SNRs after each recursion.
} 
\begin{figure}[t]
	\centering
	\includegraphics[width=3.6in,height=1.8in]{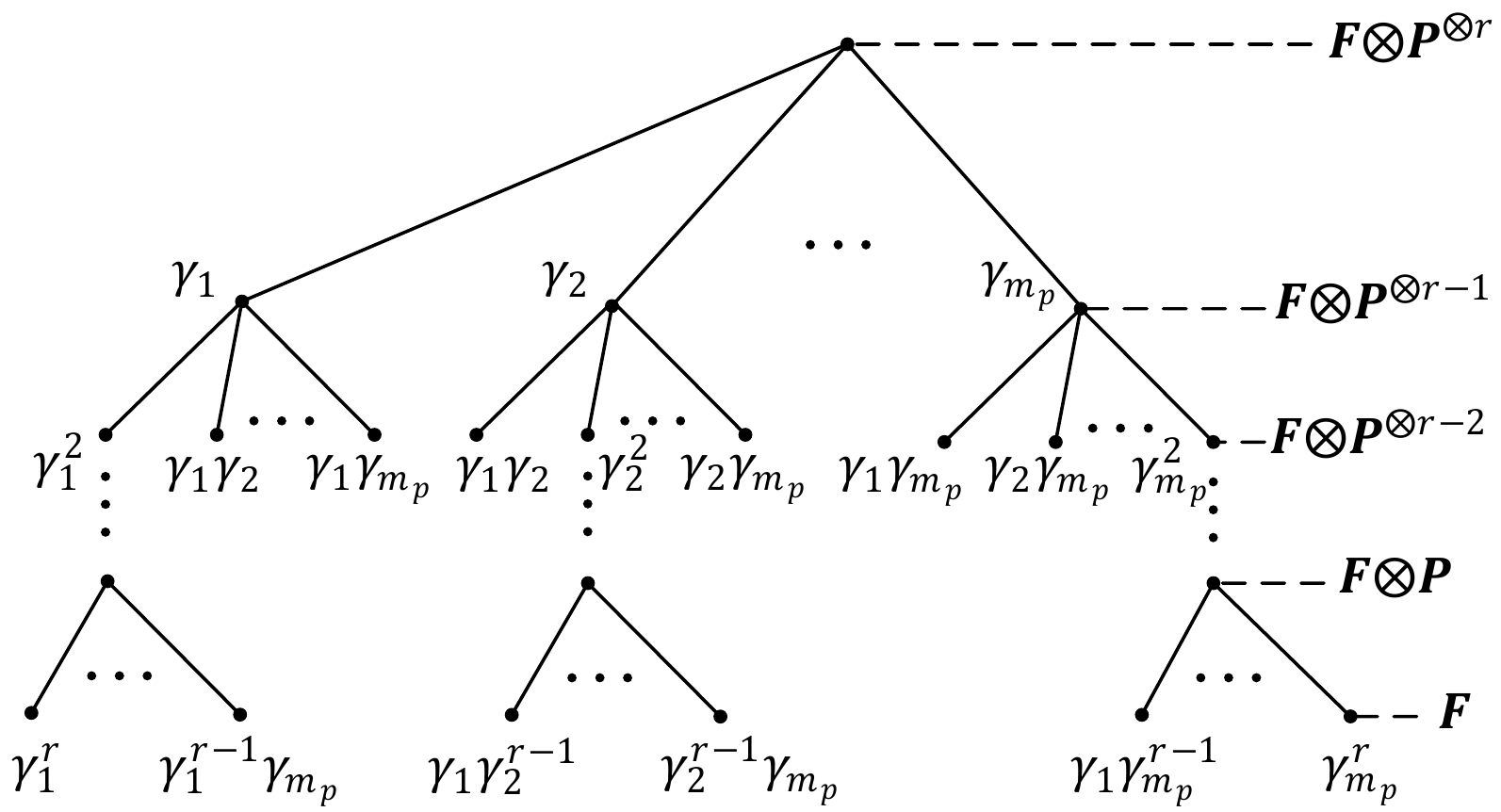}
	\caption{Tree diagram representation of the recursive detection and the resulting individual SNR gains.}
	\label{tree}
	\vspace{-0.2in}
\end{figure}

\subsubsection{Detection Complexity}
Based on the above comprehensive arguments, we can now characterize the detection complexity 
of the proposed recursive algorithm to make sure its applicability. Note that our recursive procedure results in a relatively small size for each factor matrix, even for a large $M$ and $K$; hence, it is more informative to obtain the bounds on the number of required operations instead of their orders which assumes the dimension of matrices grow large.

\noindent{\textbf{Proposition 1.}} For the proposed pattern matrix structure
$\boldsymbol{G}=\boldsymbol{F}\otimes \boldsymbol{P}^{\otimes r}$, the total number of additions/subtractions and multiplications required for the detection are at most
\begin{align}
\mathcal{N}^{\rm rec}_{\rm add}(m_f,m_p,r)&=rm_fm_p^r(m_p-1)+m_p^r\mathcal{N}^{\rm reg}_{\rm add}(\boldsymbol{F}_{m_f \times k_f}),\nonumber\\
\mathcal{N}^{\rm rec}_{\rm mul}(m_f,m_p,r)&=m_p^r\mathcal{N}^{\rm reg}_{\rm mul}(\boldsymbol{F}_{m_f \times k_f}),
\end{align}
respectively, where $\mathcal{N}^{\rm reg}_{\rm add}(\boldsymbol{F}_{m_f \times k_f})$ and $\mathcal{N}^{\rm reg}_{\rm mul}(\boldsymbol{F}_{m_f \times k_f})$ are the number of additions/subtractions and multiplications in a regular PDMA system \cite{chen2017pattern} that can be obtained given the constellation size, characteristics of the pattern matrix $\boldsymbol{F}_{m_f \times k_f}$, and the detection method applied to $\boldsymbol{F}$ such as MAP, ML, and MPA. 

{\textit{Proof:}} Based on the detection procedure in Section IV-B1, for the pattern matrix structure $\boldsymbol{G}_{M \times K}=\boldsymbol{F}_{m_f \times k_f}\otimes \boldsymbol{P}^{\otimes r}_{m_p \times m_p}$, we will end up with $M=m_fm_p^r$ equations representing the received signals over each of $M$ REs. By applying our recursive combining method to these $M$ equations we can convert them to $m_p^r$ set of $m_f$ equations, each containing at most $k_f$ auxiliary data symbols, defined based on the rectangular factor matrix $\boldsymbol{F}$. In order to obtain each of these $M$ new equations, we need to combine at most $m_p$ equations, using $m_p-1$ additions/subtractions, at each of $r$ recursions. Therefore, these new $m_p^r$ sets of $m_f$ equations can be obtained using only at most {$rm_fm_p^r(m_p-1)$} additions/subtractions. Then we can apply any advanced detection method to the resulting $m_p^r$ sets of equations which requires an additional $m_p^r$ times the number of operations over a much smaller factor matrix $\boldsymbol{F}$.

\noindent{\textbf{Example 2.}} A PDMA system with the considered parameters in Example 1 needs only $9$ times the required operations for applying any advanced detection algorithm to a regular PDMA system with a much smaller pattern matrix $\boldsymbol{F}=\begin{bmatrix}
1 & 1
\end{bmatrix}$, and an extra $36$ additions/subtractions.

\noindent{\textbf{Remark 4.}} The proposed method can also significantly reduce the latency because it enables parallel processing for the multiuser detection at the receiver side. In fact, as shown in Example 1 and Fig. \ref{tree}, we can apply the recursive detection to simultaneously process the data of different users at each recursion, i.e., concurrently process {$m_p^{l-1}$} sets of equations {(super-groups)} at each $l$-th recursion, $l=1,2,..,r$. It is worth noting that for the regular PDMA design, the rather simple SIC detection needs to process the data of each user one-by-one in a serial fashion.
\subsection{Average Sum Rate}
Based on the discussions in the previous parts and the system model in this paper, we can obtain the average sum rate of the system using the following theorem.
 
\noindent{\textbf{Theorem 2.}} The per-RE average sum rate of code-domain NOMA with the pattern matrix $\boldsymbol{G}_{M \times K}=\boldsymbol{F}_{m_f \times k_f}\otimes \boldsymbol{P}^{\otimes r}_{m_p \times m_p}$ and the proposed recursive detection can be expressed as 
\vspace{-0.05in}
\begin{align}\label{eq14}
C_M=\frac{1}{2M}&\sum_{r_1+r_2+...+r_{m_p}=r}~\frac{r!}{r_1!r_2!...r_{m_p}!}\nonumber\\
&\times\log_2\det\left({\boldsymbol{I}}_{m_f}+\mathsf{snr}\gamma_1^{r_1}\gamma_2^{r_2}...\gamma_{m_p}^{r_{m_p}}\boldsymbol{F}\boldsymbol{F}^t\right),
\end{align}
where ${\boldsymbol{I}}_{m_f}$ is an $m_f\times m_f$ identity matrix, $\boldsymbol{F}^t$ is the transpose of $\boldsymbol{F}$, and $\mathsf{snr}=P_x/\sigma^2$ in which $P_x=\E\left[x^2_j\right]$, $j=1,2,...,K$, and $\sigma^2=\E\left[n^2_i\right]$, $i=1,2,...,M$.

{\textit{Proof Sketch:}} It is well known that for a regular PDMA with the pattern matrix $\boldsymbol{A}_{M\times K}$ and the average SNR of $\mathsf{snr}$ for all of the received original data symbols, the per-RE average sum rate is given by $C^{\rm PDMA}_M=\frac{1}{2M}\log_2\det\left({\boldsymbol{I}}_{M}+\mathsf{snr}\boldsymbol{A}\boldsymbol{A}^t\right)$ for the optimal MAP detection \cite{shental2017low}. Using the recursive detection, as elaborated in Section IV-B1 and schematically shown using a tree diagram in Fig. \ref{tree}, each of $r$ square matrices manifest their effect as an increase on the SNR and finally we end up with a rectangular factor matrix $\boldsymbol{F}$.
 Then it is easy to verify using the tree digram that square matrices increase the average SNR as a factor of $\gamma_1^{r_1}\gamma_2^{r_2}...\gamma_{m_p}^{r_{m_p}}$ such that $r_1+r_2+...+r_{m_p}=r$.

\noindent{\textbf{Remark 5.}} Note that $\sum_{\Lambda_r}~\frac{r!}{r_1!r_2!...r_{m_p}!}=m_p^r$, where $\Lambda_r=\{r_1,r_2,...,r_{m_p}:\sum_{i'=1}^{m_p}r_{i'}=r\}$. Therefore, \eqref{eq14} contains the summation of $m^r_p$ REs embedded in the Kronecker product of $\boldsymbol{P}^{\otimes r}$, and the remaining portion of $m_f=M/m_p^r$ will be spanned by the rectangular matrix $\boldsymbol{F}$ in \eqref{eq14}.

\noindent{\textbf{Example 3.}} For the system parameters in Example 1 the average sum rate of all users per RE in \eqref{eq14} simplifies to $C_9=0.5\log_2\left(1+2(4/3)^2\mathsf{snr}\right)$, which can also be inferred from \eqref{eq11}.

\noindent{\textbf{Remark 6; Optimal Factor Matrices.}} When the design target is to maximize the average sum rate, the \textit{optimal pattern matrix} design can be formulated as the selection of an appropriate rectangular matrix $\boldsymbol{F}^*$ together with the $v^*$-th square matrix in Algorithm 1 with the corresponding set of SNRs $\{\gamma_{v^*,1},\gamma_{v^*,2},...,\gamma_{v^*,m_p}\}$, such that they result in the maximum sum rate in \eqref{eq14}. Satisfying further design targets such as the users' individual rate \cite{xu2015new} deserves future studies.

\noindent{\textbf{Remark 7.} Given the $v$-th square factor matrix $\boldsymbol{P}_v$, we can apply SIC detection to some of the $m_p$ equations in \eqref{eq13} to detect the $j$-th auxiliary data symbol $x_j^l$ with a higher SNR $\gamma_{v,j}$, for some $j\in\{1,2,...,m_p\}$. This evidently increases the average sum rate in \eqref{eq14}; however, it may increase the error probability due to the error propagation problem in SIC detection, resulting in an interesting trade-off between the rate and reliability. This mechanism also increases the latency because we should wait for the detection of some symbols before starting the detection of the others. 

\noindent{\textbf{Example 4.}} After detecting the auxiliary data symbols $t_1$ and $t_4$ in \eqref{eq9} using the first and second equations in \eqref{eq10}, we can form $Z_4+Z_7-t_1-t_4=4t_7+n'_4+n'_7$, instead of the third equation of \eqref{eq10}, to detect $t_7$ with a twice larger SNR rather than the factor of $4/3$. In this case, using \eqref{eq14}, the average sum rate will increase to $C_9=(4/18)\log_2\left(1+2(4/3)^2\mathsf{snr}\right)+(4/18)\log_2\left(1+2 (8/3)\mathsf{snr}\right)+(1/18)\log_2\left(1+2(2)^2\mathsf{snr}\right)$.
 
\section{Numerical Results}
In this section, we provide the numerical results for the average sum rate of various configurations, including the traditional OMA, regular PDMA with the reported optimal pattern matrices in \cite{chen2017pattern} such as $3\times 6$ and $4\times 8$ matrices with the row weight $d_f=4$, and our proposed recursive code-domain NOMA with the square factor matrices obtained from Algorithm 1. By running Algorithm 1 for $m_p=3$, we get the optimal square matrix $\boldsymbol{G}_{3\times 3}$ in Eq. \eqref{eq7} and the combining coefficients and individual SNRs reported in Example 1. Moreover, for $m_p=4$ Algorithm 1 results in four optimal square matrices of the same form and the same set of SNRs. A sample of such an optimal square matrix $\boldsymbol{P}_{4\times 4}$ along with the corresponding combining matrix $\boldsymbol{\alpha}_{4\times 4}$ is as follows
\vspace{-0.08in}
\begin{align}\label{eq15}
\!\boldsymbol{P}_{4\times 4}\!=\!
\begin{bmatrix}
0 & 0 & 0 & 1 \\
0 & 1 & 1 & 0\\ 
1 & 0 & 1  & 0\\
1 & 1 & 0 & 0
\end{bmatrix}\!,~
\boldsymbol{\alpha}_{4\times 4}\!=\!
\begin{bmatrix}
0 & -1 & 1 & 1 \\
0 & 1 & -1 & 1\\ 
0 & 1 & 1  & -1\\
1 & 0 & 0 & 0
\end{bmatrix}.\!
\end{align}
Therefore, based on the combinations defined by the matrix product of $\boldsymbol{\alpha}\boldsymbol{P}$, we get the SNR gains of $\gamma_1=\gamma_2=\gamma_3=4/3$, and $\gamma_4=1$.

\begin{figure}[t]
 \centering
 \includegraphics[width=3.6in]{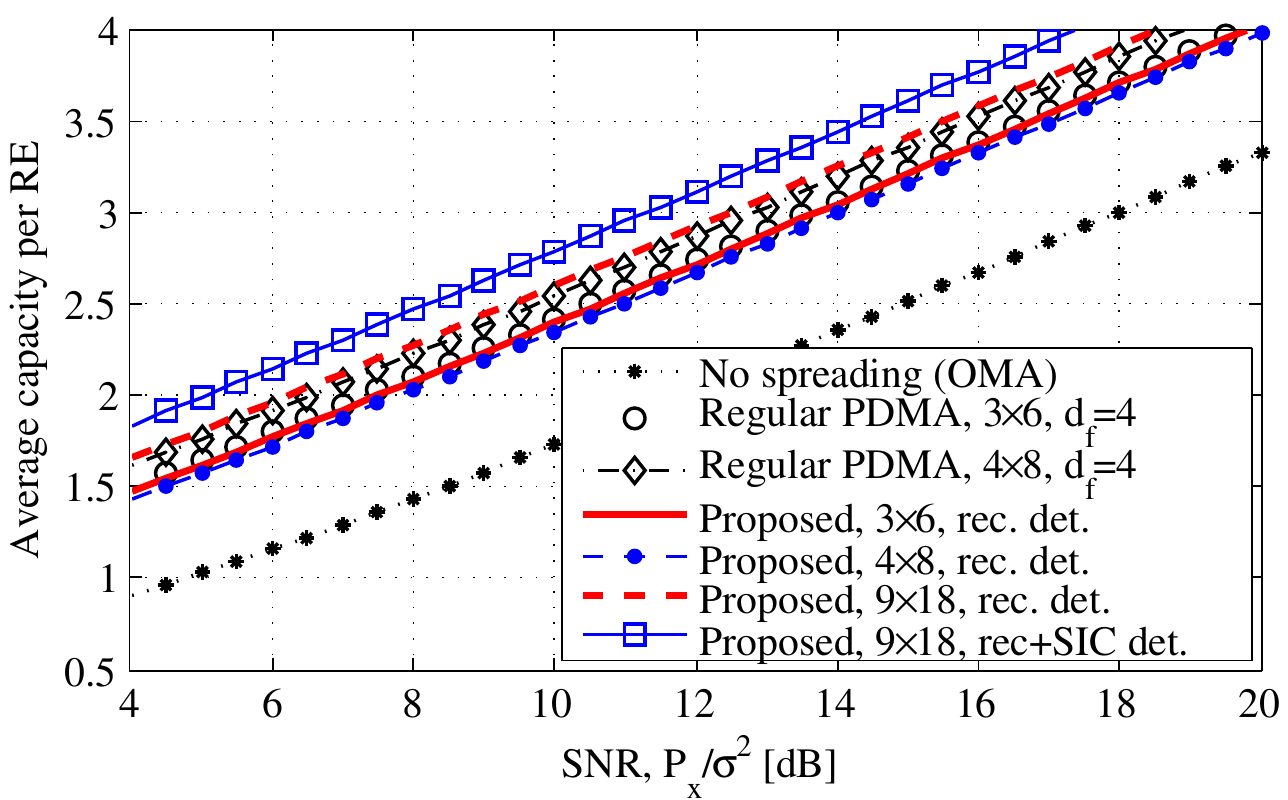}
 \caption{Average capacity of various multiple access mechanisms including the traditional OMA, regular PDMA, and our proposed recursive method.}
 \label{fig2}
 \vspace{-0.2in}
 \end{figure}
 Fig. \ref{fig2} shows the per-RE average sum rate of various multiple access techniques. It is observed that both regular PDMA and our proposed code-domain NOMA mechanisms significantly outperform the traditional OMA method. For the simulation of our proposed method we obtained the square factor matrices from Algorithm 1 and assumed the rectangular factor matrix $\boldsymbol{F}=\begin{bmatrix}
 1 & 1
 \end{bmatrix}$. Fig. \ref{fig2} demonstrates that, for relatively similar system parameters, our proposed method with the simple recursive detection approaches the capacity of regular PDMA with the optimal pattern matrix $\boldsymbol{A}_{M\times K}$ and optimal MAP detection, i.e., $C^{\rm PDMA}_M=\frac{1}{2M}\log_2\det\left({\boldsymbol{I}}_{M}+\mathsf{snr}\boldsymbol{A}\boldsymbol{A}^t\right)$. However, due to the remarkably lower complexity of our method, we can exploit higher pattern matrix dimensions (by increasing $r$) and also apply SIC detection at some recursions similar to Example 3 to get larger sum rates while achieving such gains through larger pattern matrix dimensions are questionable for the regular PDMA.
\section{Conclusion}
In this paper, we proposed a low-complexity recursive approach toward code-domain NOMA based on factorizing the spreading pattern matrix as the Kronecker product of some smaller factor matrices. We observed that the proposed method, with the simple recursive detection, approaches the average sum rate of regular PDMA technique with optimal detection. The proposed method benefits from a significantly lower complexity and latency and, through realizing large-dimension pattern matrices, paves the way toward massive nonorthogonal multiplexing for the broad range of 5G applications. 
\end{document}